\documentstyle[12pt]{article}
\def\strut{\rule[-.5cm]{0cm}{1cm}}
\def\dspace{\baselineskip = .30in}

\def\VEV#1{\left\langle #1\right\rangle}
\begin{document}

\title{Natural Suppression of Higgsino-Mediated Proton Decay in
Supersymmetric SO(10)\thanks{Supported in part by Department of Energy
Grant \#DE-FG02-91ER406267}}

\author{{\bf K.S. Babu} and {\bf S.M. Barr}\\
Bartol Research Institute\\
University of Delaware\\ Newark, Delaware
19716}

\date{BA-93-26, May 1993}
\maketitle
\begin{abstract}

In supersymmetric Grand Unified Theories, proton decay mediated by the
color--triplet higgsino is generally problematic and requires
some fine--tuning
of parameters.  We present a mechanism which naturally
suppresses such dimension 5 operators in the context of SUSY $SO(10)$.
The mechanism, which  implements natural doublet--triplet splitting
using the adjoint higgs, converts these dimension 5 operators
effectively into
dimension 6.  By explicitly computing the higgs
spectrum and the resulting threshold uncertainties we show that
the successful prediction of
$\sin^2\theta_W$ is maintained {\it as a prediction} in this
scheme.  It is argued that only a weak suppression of the higgsino
mediated proton decay is achievable within SUSY $SU(5)$ without
fine--tuning, in contrast to a strong suppression in
SUSY $SO(10)$.
\end{abstract}
\newpage

\dspace
\section*{1. Introduction}
The dramatically precise unification of couplings$^1$ that occurs in
the minimal supersymmetric $SU(5)$ model has been much touted, and
indeed is striking. A fit$^2$ to all W,Z, and neutral current data
(using $m_t = 138 GeV$ and $m_H = M_Z)$ gives $\sin^2\theta_W(M_Z) =
0.2324 \pm 0.0003$, while in SUSY $SU(5)$ one has$^2$
$\sin^2\theta_W(M_Z) = 0.2334 \pm 0.0036$ (we have combined the
uncertainties due to $\alpha_s(M_Z),\; \alpha(M_Z)$, sparticle
thresholds, $m_t, m_{h_{0}}$, high-scale thresholds and
non-renormalizable operators). As a measure of how significant this
agreement is consider that the addition of just one extra pair of
light higgs doublet supermultiplets would increase the SUSY GUT
prediction of $\sin^2\theta_W(M_Z)$ to about 0.256, leading to a
discrepancy of over six standard deviations.

At the same time SUSY $SU(5)$ suffers from a problem$^3$ with proton
decay arising from dimension-5 operators caused by the exchange of
color-triplet higgsinos in the same SUSY-$SU(5)$ multiplet with the
light higgs doublets, $H$ and $H^\prime$. For `central values' of the
model parameters the predicted proton life-time mediated by the
dimension-5 operators$^{4,5}$ is shorter than the current experimental
limits.$^6$  A certain amount of adjustment is then required for
consistency, which pushes parameters to the corner of their allowed
region. For short we will henceforth refer to this as the
`higgsino-mediated proton decay (HMPD) problem'.

It would seem that one cannot take seriously the unification of
couplings of SUSY-GUTs, however impressive, in the absence of
satisfactory mechanism that suppresses higgsino-mediated proton
decay. Two requirements for a `satisfactory' mechanism that we will
impose are that it involves no `fine-tuning' or artificial
adjustments of parameters, and that the unification of couplings is
maintained as a \underline{prediction}. (We emphasize that word
because a discrepancy in $\sin^2\theta_W$ can often be remedied by
introducing \underline{ad} \underline{hoc} new particles, threshold
effects, etc; but we would not regard the resulting agreement as
being in any sense a prediction.)

The higgsino-mediated proton decay problem$^3$ is easily described.
In SUSY $SU(5)$ models there exists a pair of higgs super-multiplets,
that we will denote $5_H + \bar{5}_H^\prime$. Under the standard
model group, $G_S = SU(3) \times SU(2) \times U(1)$, these decompose
as $5_H = \{(1,2,+\frac{1}{2}) + (3,1,-\frac{1}{3})\} \equiv \{2_H +
3_H\}$, and $\bar{5}_H^\prime = \{(1,2,-\frac{1}{2}) +
(\bar{3},1,+\frac{1}{3})\} \equiv \{\bar{2}_H^\prime +
\bar{3}_H^\prime\}$. The $2_H$ and $\bar{2}_H^\prime$ are just the
familiar $H$ and $H^\prime$ of the supersymmetric standard model, and
their couplings to the light quarks and leptons are therefore fixed
in terms of their vacuum expectation values and the light fermion
masses. Then by $SU(5)$ the Yukawa couplings of the $3_H$ and
$\bar{3}_H^\prime$ are fixed as well. If there is a Dirac mass term
connecting the higgsinos in $3_H$ and $\bar{3}_H^\prime$ to each
other, i.e., a term of the form $M (3_H \bar{3}_H^\prime)$, then the
diagram shown in figure 1 exists, which depicts a baryon-number
violating process mediated by colored higgsino exchange. The
higgsino, of course, converts the quarks and leptons to their scalar
partners, so figure 1 needs to be `dressed' for it to correspond to
proton decay. Dressing by $W$-ino exchange is by far the most
dominant, the resulting life-time for $p \rightarrow
K^+\bar{\nu}_\mu$ for example (the anti-symmetry of the relevant
operator causes the change in flavor) has been estimated$^4$ to be

\begin{equation}\begin{array}{lcl}
\tau(p \rightarrow K^+\bar{\nu}_\mu) & = & 6.9 \times 10^{28} yr \times
\left| \left( \frac{0.01 GeV^3}{\beta}\right)
\left(\frac{0.67}{A_s}\right) \left( \frac{\sin 2\beta_H}{1 +
y^{tk}}\right)\right.\strut\\
& & \left.\left( \frac{M_{Hc}}{10^{16} GeV} \right)
\frac{TeV^{-1}}{f(u,d) + f(u,e)} \right|^2.\end{array}
\end{equation}

\noindent
Here $\beta$ is the relevant nuclear matrix element which lies in the
range $\beta = (0.003 - 0.03) GeV^3$. $A_s$ is the short-distance
renormalization factor $(A_s \simeq 0.6),\; \tan\beta_H =
\VEV{H}/\VEV{H^\prime}$, and $y^{tk}$ parametrizes the contribution of the
top family relative to the first two $(0.1 < \mid y^{tk} \mid < 1.3$
for $m_t = 100 GeV)$. The functions $f$ arise from the one-loop
integrals and $f(u,d) \simeq
m_{\tilde{W}^\pm}/m^2_{\tilde{Q}}$ for
$m_{\tilde{W}^\pm} \ll m_{\tilde{Q}}$.

The prediction (1) is to be compared with the present experimental
lower limit $\tau (p \rightarrow K^+\bar{\nu}_\mu)
\stackrel{_>}{_\sim} 1 \times 10^{32}$ yr.$^6$ There is already, as
can be seen, somewhat of a difficulty reconciling these numbers. At
least several of the following conditions should be fulfilled: (a)
The nuclear matrix element $\beta$ is near the lower end, $\beta
\simeq 0.003 ~GeV^3$, (b) $\tan\beta_H$ is not too large, (c) either
the $\tilde{W}^\pm$ is significantly lighter than the squark
$\tilde{Q}$ or vice versa, (d) the colored higgsino mass should
exceed the GUT scale significantly $(M_{Hc} \stackrel{_>}{_\sim}
10^{17} GeV)$, (e) there is some cancellation in $(1 + y^{tk})$
between the third family and the first two family contributions.
Obviously this pushes almost all parameters to their corners.
Although not excluded, the problem begs for a more elegant
explanation.

The necessity of the Dirac mass term $M (3_H \bar{3}_H^\prime)$ for
obtaining the $D = 5$ baryon-violating operator is a crucial point. A
quartic term in the chiral superfields will be suppressed by
$\frac{1}{M_{GUT}}$ or $\frac{1}{M_{GUT}^2}$ depending on whether it
is an $F$-term or a $D$-term. To get an $F$-term all the left-handed
chiral superfields must be coming into the diagram (or out of it).
This requires the chirality-flipping mass insertion coming from the
$M (3_H \bar{3}_H^\prime)$ term as shown in fig. 1. If the mass $M$
vanished, one could only write superfield diagrams like fig. 2, which
clearly give $D$-terms and therefore are suppressed by
$\frac{1}{M_{GUT}^2}$. (The $F$-terms correspond to higgsino
exchange, the $D$-terms to higgs exchange.)

The foregoing considerations have suggested to several
authors$^{7,8}$ an approach to suppressing higgsino-mediated proton
decay by imposing a (Peccei-Quinn type) symmetry that suppresses the
Dirac mass between the $3_H$ and $\bar{3}_H^\prime$. However, this
approach leads to a dilemma. In the simplest SUSY $SU(5)$ model it is
precisely the $M (3_H \bar{3}_H^\prime)$ Dirac term that gives to the
$3_H$ and $\bar{3 ^\prime}_H$ the superlarge mass that they must have (else
even the $D = 6$ operators would cause a disaster). How, then, can
the $3_H$ and $\bar{3}_H^\prime$ be made superheavy and yet not have a
large Dirac mass connecting them to each other? To resolve this, one
ends up introducing new color-triplet higgs superfields, a
$\bar{3}_H$ to mate with the $3_H$ and a $3_H^\prime$ to mate with
the $\bar{3}_H^\prime$, so that all color triplets become heavy while
still leaving the unprimed and primed sectors disconnected. That
means having four instead of the minimal two 5-plets of higgs. The
situation can be represented diagrammatically as follows

\begin{equation}\begin{array}{cccc}
\left( \begin{array}{c}
3\\2 \end{array} \right) & \left(       \begin{array}{c}\bar{3}\\\bar{2}
\end{array} \right) & \left(
\begin{array}{c}3^\prime\\2^\prime\end{array} \right) & \left( \begin{array}{c}
\bar{3}^\prime\\\bar{2}^\prime\end{array} \right)\strut\\
\parallel & \parallel & \parallel & \parallel\strut\\
5_H & \bar{5}_H & 5_H^\prime & \bar{5}_H^\prime\end{array}
\end{equation}

\noindent
where the solid horizontal lines representing superlarge Dirac
masses. The $2$ and $\bar{2}^\prime$ are the usual light $H$ and
$H^\prime$ of the MSSM. But now it is to be noticed that there are
two additional light doublets, $\bar{2}$ and $2^\prime$. As noted
above, this is unacceptable if the dramatic unification of gauge
couplings is to be preserved as a prediction.

One could remove the extra pair of Higgs(ino) doublets to superlarge
scales by introducing a mass term $M (\bar{5}_H 5_H^\prime)$, indicated
by the dashed line in eq. (2). This, of course, would reintroduce the
higgsino-mediated proton decay as shown in fig. 3. The parameter
$M$ would then control both the higgsino-mediated proton decay and
the mass of the pair of extra doublets.

As we shall see in the later sections, this situation is typical of
mechanisms that naturally suppress higgsino-mediated proton decay:
(a) there is a doubling of the higgs sector, (b) there is the
consequent danger to the unification of couplings of extra light
fields in incomplete multiplets, and (c) there is a parameter $M$
which controls both proton decay and the mass of these extra fields.

Two approaches, therefore, appear to be possible. (1) \underline{Weak
suppression} of higgsino-mediated proton decay, resulting from $M$
being of order -- but slightly less than -- $M_{GUT}$. For example,
with $M = \frac{1}{10} M_{GUT}$ higgsino-mediated proton decay is
suppressed by a factor of $10^{-2}$ while at the same time those
extra fields whose mass is given by $M$ will only lead to small
threshold corrections to $\sin^2\theta_W$. In this case the
suppression of higgsino-mediated proton decay is just numerical;
there is no symmetry or other qualitative explanation of it. One
would have no \underline{a priori} reason therefore to
expect the suppression to be particularly large. A hope would
therefore exist that $p \rightarrow K^+\bar{\nu}_\mu, n \rightarrow
K^\circ\bar{\nu}$, etc. might be seen at super-Kamiokande. (2)
\underline{Strong suppression} of higgsino-mediated proton decay
would result if (due to some approximate symmetry perhaps) $M$ were
much less than $M_{GUT}$, say, $O(M_W)$. In that case it is
imperative that there be no `extra' fields (i.e., beyond the minimal
supersymmetric standard model) in incomplete $SU(5)$ multiplets whose
mass is proportional to $M$, or else the unification of couplings
would be destroyed. To achieve this without fine-tuning turns out to
be a non-trivial problem. One of the main conclusions of this paper
is that such a strong suppression of proton decay can only be
achieved in a satisfactory way in $SO(10)$ (or larger groups).

The whole problem of higgsino-mediated proton decay is of course
intimately connected to the well known question of `doublet-triplet'
splitting.$^{9-13}$ We will show that the most satisfactory
treatments of this problem make use of an old but somewhat neglected
idea for doublet-triplet splitting in $SO(10)$ using the
adjoint higgs due to Dimopoulos and
Wilczek.$^{13}$

Our paper is organized as follows. In section II we review the
Dimopoulos-Wilczek idea for doublet-triplet splitting and show how
both weak suppression and strong suppression of HMPD can be achieved
naturally in SUSY $SO(10)$. In section III we consider SUSY $SU(5)$ and
show that only weak suppression of HMPD can be achieved without
fine-tuning parameters. In section IV we discuss flipped $SU(5)$. In
section V a closer examination of the Dimopoulos-Wilczek mechanism in
$SO(10)$ is undertaken, and we show that it can be made viable and
consistent. Our conclusions are summarized in section VI. In appendix
A we give the details of the minimization of a realistic $SO(10)$
superpotential.  There we show that the gauge symmetry breaking can be
achieved consistent with supersymmetry without generating pseudo
goldstone bosons.
Appendix B deals with the threshold corrections to
$\sin^2\theta_W$.

\section*{2. Suppressing Proton Decay in SUSY SO(10)}

The problem with doublet-triplet splitting arises in $SU(5)$ because of
the tracelessness of its irreducible representations. Suppose the
superpotential contains the term $\lambda_1 \bar{5}_H^\prime 24_H
5_H$ and $\VEV{24_H} = {\rm diag} (x, x, x, y, y)$, then the $3_H$ and
$\bar{3}_H^\prime$ higgsinos get a Dirac mass of $\lambda_1 x$ and
the doublets $2_H$ and $\bar{2}_H^\prime$ get a mass $\lambda_1 y$.
We need $\lambda_1 x \sim M_{GUT}$ and $\lambda_1 y \sim M_W$ but
this is impossible since, by the tracelessness of $24$, $y = -
\frac{3}{2} x$. This can be remedied by introducing a singlet
superfield,
$1_H$, with the coupling $\lambda_2 \bar{5}_H^\prime 1_H
5_H$ and $\VEV{1_H} \equiv z$ (or equivalently by a bare mass term),
but only by tuning the parameters so that
$(- \frac{3}{2} \lambda_1 x + \lambda_2 z) \stackrel{_<}{\sim}
10^{-14} (\lambda_1 x + \lambda_2 z)$.

In $SO(10)$ such fine adjustment of parameters can be avoided because
the analogue of the tracelessness condition does not exist.$^{13}$  The
24, which is the adjoint of $SU(5)$, is contained in the 45 which is the
adjoint of $SO(10)$. 45 is a rank-2 antisymmetric tensor and the VEV of
$45_H$ can be brought to the canonical form

\begin{equation}\begin{array}{ccl}
\VEV{45_H} & = & \eta \otimes {\rm diag} (x_1, x_2, x_3, x_4, x_5),\strut\\
\eta & \equiv & \left( \begin{array}{cc} 0\\-1\end{array}
\begin{array}{cc} 1\\0\end{array} \right),\end{array}
\end{equation}

\noindent which corresponds to the $U(5)$ matrix ${\rm diag} (x_1, x_2,
x_3, x_4, x_5)$. Because this is a $U(5)$ rather than an $SU(5)$
matrix its trace need not vanish. One can therefore have the VEV of
$\VEV{45_H}$ take the form

\begin{equation}
\VEV{45_H} = \eta \otimes {\rm diag} (a, a, a, 0, 0,).
\end{equation}

\noindent
This is just what is needed to give mass to the $SU(3)_C$ - triplet
higgs(inos) and not the $SU(2)_L$ - doublet ones. This is what we
call the Dimopoulos-Wilczek mechanism.

There is another group--theoretical explanation for the
doublet-triplet splitting in $SO(10)$. Under its maximal subgroup
$SU(2)_L \times SU(2)_R \times SU(4)_C$, the standard model singlets
of 45 which could acquire GUT-scale VEVs are contained in the
(1, 1, 15) and (1, 3, 1) multiplets. The  10 of Higgs decomposes
as (2, 2, 1) + (1, 1, 6). If only the (1, 1, 15) of 45 acquires
a VEV, it gives the color triplets of (1, 1, 6) a mass and not the
doublets of (2, 2, 1). If the (1, 3, 1) acquires a VEV, it will
supply a super-large mass to the doublets - and not to the triplets.
(We shall shortly make use of the second property to suppress proton
decay and at the same time preserve $\sin^2\theta_W$ as a
prediction.)  Such options are not available in $SU(5)$, since $SU(5)$ has
no intermediate symmetries, even for the sake of classification.

Consider the following coupling in the superpotential of a
SUSY-$SO(10)$ model:

\begin{equation}
W \supset \lambda 10_{1 H} 45_H 10_{2 H}~.
\end{equation}

\noindent
One must introduce \underline{two} 10's of Higgs(ino) fields because
with just one the term $10_H45_H10_H$ would vanish by the
antisymmetry of the 45. As we noted in the introduction, and shall
see more clearly later, such a doubling is actually a useful thing
from the point of view of suppressing higgsino-mediated proton decay.
This is another appealing feature of $SO(10)$.

When the $45_H$ gets the VEV shown in eq (4) all of the triplet
Higgs(ino) fields in $10_{1 H}$ and $10_{2 H}$ get superlarge masses.
The situation can be represented schematically as follows

\begin{equation}\begin{array}{cccc}
\left( \begin{array}{c}
{3}_1\\{2}_1 \end{array} \right) & \left(
\begin{array}{c}\bar{3}_2\\\bar{2}_2
\end{array} \right) & \left(
\begin{array}{c}{3}_2\\{2}_2\end{array} \right) & \left( \begin{array}{c}
\bar{3}_1\\\bar{2}_1\end{array} \right)\strut\\
\parallel & \parallel & \parallel & \parallel\strut\\
{5}_{1 H} & \bar{5}_{2 H} & {5}_{2 H} & \bar{5}_{1 H}\end{array}
\end{equation}

\noindent
where under $SO(10) \rightarrow SU(5), 10_{1 H} = \bar{5}_{1 H} +
5_{1 H}$, and $10_{2 H} = \bar{5}_{2 H} + 5_{2 H}$. By comparison
with the scheme shown in eq. (2) one sees that the `unprimed sector'
consists of ${5}_{1 H}$ and $\bar{5}_{2 H}$, while the `primed sector'
consists of ${5}_{2 H}$ and $\bar{5}_{1 H}$. One can then identify $H
\equiv 2_{1 H}$ and $H^\prime = \bar{2}_{1 H}$.

Now we face the problem of generating mass for the `extra' doublets
which reside in $5_{2 H} + \bar{5}_{2 H} = 10_{2 H}$. The simplest
possibility is just to introduce into the superpotential, $W$, the
term $M (10_{2 H} 10_{2 H})$, with $M/M_{GUT}$ being less than - but
not \underline{much} smaller than - one. The resulting threshold
correction to $\sin^2\theta_W$ is $+ \frac{3\alpha(M_Z)}{10\pi} \ell
n (M_{GUT}/M) \approx 10^{-3}$. For proton decay to be suppressed it
is also necessary that only $10_{1 H}$ but not $10_{2 H}$ couple to
light quarks and leptons with usual strength. All of this (including
the absence of a $(10_{1 H})^2$ term in $W$) can be enforced by a
global symmetry. For example, one could have a symmetry under which
$10_{2 H} \rightarrow + 10_{2 H}, 10_{1 H} \rightarrow - 10_{1 H},
45_H \rightarrow - 45_H$, and $16_j \rightarrow i16_j$ (the light
families, with $j = 1,2,3$).

The above appears to us to be the simplest way of achieving weak
suppression of proton decay. (For comparison with $SU(5)$ see the
next section.) However, in $SO(10)$, but not in $SU(5)$, it is
actually possible to achieve a \underline{strong suppression} in a
satisfactory way. To do this we need to give $2_{2H}$ and $\bar{2}_{2H}$ a
superheavy Dirac mass (so as to not mess up $\sin^2\theta_W$) without
having a superheavy Dirac mass connecting $3_{2H}$ and $\bar{3}_{2H}$
(which could produce excessive proton decay). But this is
just a doublet-triplet splitting problem -- but upside down to the
familiar one! Here the doublets but not the triplets need a mass
term. This will prove to be not doable in $SU(5)$ without tuning, but
in $SO(10)$ it can be done. What is required is another $45_H$ with a
VEV

\begin{equation}
\VEV{45_H^\prime} = \eta \otimes {\rm diag} (0, 0, 0, a^\prime, a^\prime)
{}~.
\end{equation}

\noindent
As already noted, this is just as achievable as the VEV given in eq.
(4). (See section V for the demonstration.)

There is a slight hitch in that we cannot, because of the
antisymmetry of $45_H^\prime$, simply write down $10_{2 H}
45_H^\prime 10_{2 H}$ to give mass to $2_{2 H}$ and $\bar{2}_{2 H}$.
However, this can be overcome by introducing an additional 10 of
Higgs(ino) fields. Consider a superpotential containing

\begin{equation}\begin{array}{lcl}
W & \supset & \lambda 10_{1 H} 45_H 10_{2 H} + \lambda^\prime 10_{2 H}
45_H^\prime 10_{3 H}\strut\\
& + & M 10_{3 H} 10_{3 H}\strut\\
& + & \sum_{i,j=1}^3\; f_{ij} 16_i 16_j 10_{1 H}\end{array}
\end{equation}

\noindent
with $\VEV{45_H}$ and $\VEV{45_H^\prime}$ being given by eqs. (4) and (7).
Then the superheavy mass matrices of the color-triplet and
weak-doublet higgs(inos) are

\begin{equation}
\left( \bar{2}_1, ~\bar{2}_2,~ \bar{2}_3 \right) \left( \begin{array}{ccc}
0 & 0 & 0\strut\\ 0 & 0 & \lambda^\prime a^\prime\strut\\ 0 &
-\lambda^\prime a^\prime & M\end{array} \right) \left( \begin{array}{c}
2_1\strut\\ 2_2\strut\\ 2_3\end{array} \right)
\end{equation}

\noindent
and

\begin{equation}
\left( \bar{3}_1,~ \bar{3}_2,~ \bar{3}_3 \right) \left(
\begin{array}{ccc} 0 & \lambda a & 0\strut\\ -\lambda a & 0 &
0\strut\\ 0 & 0 & M\end{array} \right) \left( \begin{array}{c}
3_1\strut\\ 3_2\strut\\ 3_3\end{array} \right)~.
\end{equation}

\noindent
The doublet matrix is rank-two leaving a single pair of light
doublets $H \equiv 2_{1 H}$ and $H^\prime \equiv \bar{2}_{1 H}$. All
triplets get superheavy mass; however, there is no mixing between
$3_{1 H}$ and $\bar{3}_{1 H}$ that would permit the diagrams shown in
figures 1 or 3.

There are several questions to be answered concerning the $SO(10)$
approaches to the proton decay problem. (1) Can the VEVs in eqs (4)
and (7) arise from an actual (super) potential?$^{14}$ (2) Can
$SO(10)$ be broken all the way to $SU(3) \times SU(2) \times U(1)$
without destabilizing these VEVs? and (3), are the threshold
corrections in such an $SO(10)$ model likely to be small enough not
to vitiate the successful prediction of $\sin^2\theta_W$? We will
show in section V that the answer to all these questions is `yes'.
But first we will examine the possibilities that exist in $SU(5)$ and
flipped $SU(5)$.

\section*{3. Suppressing Proton Decay in SUSY SU(5)}

The only viable method of doublet-triplet splitting in SUSY $SU(5)$
that does not involve fine tuning of parameters is the `missing
partner mechanism'.$^{9,10}$ The so-called sliding-singlet
mechanism$^{11}$ has the problem in $SU(5)$ that radiative
corrections destroy the gauge hierarchy.$^{12}$ For the missing
partner mechanism in
$SU(5)$ one requires (at least) the set $5 + \bar{5} + 50 + \overline{50}
+ 75$ of Higgs supermultiplets. In the $50 (\overline{50})$ there is a
color $3 (\bar{3})$ but no weak $2(\bar{2})$. Thus the couplings
$\lambda 5_H \overline{50}_H \VEV{75_H} + \lambda^\prime \bar{5}_H 50_H
\VEV{75_H}$ give mass to the triplets in $5_H + \bar{5}_H$ but not to the
doublets. Schematically,

\begin{equation}\begin{array}{cccc}
\left( \begin{array}{c} 3\strut\\ 2\end{array} \right) & \left(
\begin{array}{c} \bar{3}\strut\\ {\rm other}\end{array} \right) &
\left( \begin{array}{c} 3\strut\\ {\rm other}\end{array} \right) &
\left( \begin{array}{c} \bar{3}\strut\\ \bar{2}\end{array}
\right)\strut\\
\parallel & \parallel & &\strut\\
5_H & \overline{50}_H & 50_H & \bar{5}_H\end{array}
\end{equation}

\noindent
The horizontal solid lines represent superheavy triplet-higgs(ino)
masses arising from the $\VEV{75_H}$. As in the cases considered in
previous sections, there is the question of how to make the `other'
fields in the $50 + \overline{50}$ superheavy. (They contribute to the RGE
at one loop the same as a pair of weak doublets.) If one wanted a
`strong suppression' of proton decay, it would require giving
superlarge masses to all the `other' fields in $50 + \overline{50}$ but
having the $\bar{3} (\overline{50}_H)$ and $3(50_H)$ not be connected by a
large Dirac mass term. There is no analogue of the missing partner
mechanism that would accomplish this in a natural way. It could only
be done by fine-tuning. For example, two different representations
$(1_H$ and $24_H$, or $24_H$ and $75_H$) could couple 50 to
$\overline{50}$ and be relatively adjusted to give light mass only to the
$3 (50_H) + \bar{3} (\overline{50}_H)$. However, we have foresworn
fine-tuning.

It \underline{is} possible to achieve a natural weak suppression of
higgsino-mediated proton decay in $SU(5)$ by introducing an explicit
mass term $M (\overline{50}_H 50_H)$ into the superpotential and having
$M/M_{GUT}$ be of order but somewhat smaller than unity. This works
as well as the weak suppression mechanism in $SO(10)$ discussed in
section II. However, in $SU(5)$ there is the necessity of introducing
the somewhat exotic high rank representations $50,~\overline{50}$ and $75$,
whereas in $SO(10)$ only the low rank representations
10, 45, and 54 are
required. If one were willing to live with multiple fine-tunings of
parameters one could do with just doubling the higgs sector in
$SU(5)$ to $5 + \bar{5} + 5^\prime + \bar{5}^\prime$ as discussed
briefly in the introduction.  With \underline{two} fine-tunings one
could make all the triplets heavy, and achieve weak suppression of
proton decay. With a \underline{third} fine-tuning one could give
mass to the extra pair of doublets and yet achieve strong
suppression of proton decay. There are papers in the literature that
take this approach.$^{7,8}$

\section*{4. Suppressing Proton Decay in Flipped SU(5)}

As is well-known, the missing partner mechanism works much more
economically in flipped $SU(5)^{15}$ than in ordinary $SU(5).^{16}$ One
requires for the mechanism the $SU(5) \times U(1)$ representations
$5_H^{-2} + \bar{5}_H^2 + 10_H^1 + \overline{10}_H^{-1}$, and the
superpotential couplings $\lambda 5_H^{-2} 10_H^1 10_H^1 +
\lambda^\prime \bar{5}_H^2 \overline{10}_H^{-1} \overline{10}_H^{-1}$. The
$10_H (\overline{10}_H)$ contains a color $\bar{3} (3)$ but no
color-singlet, weak-doublet components. The $10_H$ and $\overline{10}_H$
get VEVs that break $SU(5) \times U(1)$ down to $G_S$ and also mate
the triplet higgs(inos) in the $5_H + \bar{5}_H$ with these in the
$10_H + \overline{10}_H$ leaving the doublets in $5_H + \bar{5}_H$ light.
Schematically,

\begin{equation}\begin{array}{cccc}
\left( \begin{array}{c} 3\strut\\ 2\end{array} \right) & \left(
\begin{array}{c} \bar{3}\strut\\ {\rm other}\end{array} \right) &
\left( \begin{array}{c} 3\strut\\ {\rm other}\end{array} \right) &
\left( \begin{array}{c} \bar{3}\strut\\ \bar{2}\end{array}
\right)\strut\\
\parallel & \parallel & \parallel & \parallel\strut\\
5_H & 10_H & \overline{10}_H & \bar{5}_H\end{array}
\end{equation}

\noindent
Another beautiful feature of flipped $SU(5)$ is that there is no
necessity to do anything else to give mass to the `other' fields in
the $10_H + \overline{10}_H$: they are all disposed of by the (super)
Higgs mechanism! They are either eaten or become superheavy with the
gauge/gaugino particles. Thus, in flipped $SU(5)$ one can strongly
suppress higgsino-mediated proton decay without any fine-tuning and
without leaving any `extra' split multiplets lighter than $M_{GUT}$.
We found this to be impossible in ordinary $SU(5)$. However, there is
one major drawback: the group of flipped $SU(5)$ is really $SU(5)
\times U(1)$ and so real unification of gauge couplings is not
achieved. One has therefore lost, or rather never had, the
unification of gauge couplings \underline{as a prediction}.

\section*{5. A More Detailed Examination of SO(10)}

In section II certain ideas were discussed for solving the
doublet-triplet splitting problem and for suppressing
higgsino-mediated proton decay in $SO(10)$ that made essential use of
specific patterns of VEVs, in particular those shown in eqs. (4) and
(7). The question arises whether such VEVs are natural. In ref. 14
Srednicki wrote down a superpotential for a 45 and a 54 of Higgs that
has both of these forms as possible supersymmetric minima. Let us
denote the 45 and 54 by $A$ and $S$ respectively. Then the most
general $SO(10)$-invariant superpotential involving just these fields
has the form

\begin{equation}
W(A,S) = m_1 A^2 + m_2 S^2 + \lambda_1 S^3 + \lambda_2 A^2S~.
\end{equation}

\noindent
The equations for a supersymmetric minimum are

\begin{eqnarray}
0 = F_A & =& 2(m_1 + \lambda_2 S)A \nonumber \\
0 = F_S & = & (2 m_2 S + \lambda_2 A^2 + 3 \lambda_1 S^2)~.
\end{eqnarray}

\noindent
Suppose we choose $\VEV{S} = \left( \begin{array}{cc} 1 & 0\\0 &
1\end{array} \right) \otimes {\rm diag} (s, s, s, -\frac{3}{2} s,
-\frac{3}{2} s)\; (S$ is a traceless rank-two symmetric tensor of
$SO(10)$) and $\VEV{A} = \left( \begin{array}{cc} 0 & 1\\ -1 &
0\end{array} \right) \otimes {\rm diag} (a, a, a, b, b)$. Then
equation (14) gives two equations:

\begin{equation}\begin{array}{ccc}
(m_1 - \frac{3}{2} \lambda_2 s) b & = & 0,\strut\\
(m_1 + \lambda_2 s) a & = & 0.\end{array}
\end{equation}

\noindent
Either $a$ or $b$ or both must therefore vanish (if $s \neq 0)$.
There are therefore three possible solutions. (1) $b = 0, s =
-m_1/\lambda_2$; (2) $a = 0, s = \frac{2}{3} m_1/\lambda_2$; (3) $a =
b = 0$.

For doublet-triplet splitting and weak suppression of proton decay we
need only the solution (1). For strong suppression we need (at least)
two adjoints, one with VEV corresponding to solution (1) and the other
to solution (2). (See eqs. (4) and (7)). We will examine this latter
more complicated case in greater detail. If the required pattern of
VEVs can be achieved in a realistic model for that case, then
\underline{a} \underline{fortiori} the simpler requirements for weak
suppression can be achieved also. The main issues are whether the
VEVs in eqs. (4) and (7) can be achieved, the group $SO(10)$ broken
completely to $SU(3) \times SU(2) \times U(1)$, and goldstone
particles avoided. (The issue of threshold corrections to
$\sin^2\theta_W$ is dealt with in Appendix B.)

To begin with we double the superpotential shown in equation (13).
That is, we have two 45's, denoted $A$ and $A^\prime$, and two 54's,
denoted $S$ and $S^\prime$, with superpotential

\begin{equation}\begin{array}{lcl}
W (A, S; A^\prime, S^\prime) & = & m_1 A^2 + m_2S^2 + \lambda_1 S^3 +
\lambda_2 A^2 S\strut\\
& + & m_1^\prime A^{ \prime 2} + m_2^\prime S^{ \prime 2} + \lambda_1^\prime
S^{ \prime 3} + \lambda_2^\prime A^{ \prime 2} S^\prime.\end{array}
\end{equation}

\noindent
We aim to have the $(A,S)$ sector have VEVs in solution (1), and the
$(A^\prime, S^\prime)$ sector have VEVs in solution (2).

The superpotential in (16) is certainly not enough because, for one
thing, nothing determines the relative alignment of the VEVs of the
two sectors, and so there are goldstone modes corresponding to a
continuous degeneracy whereby the sectors are rotated in
$SO(10)$-space with respect to each other.

A second problem with eq. (16) is that $\VEV{A}$ and $\VEV{S}$ break
$SO(10) \rightarrow [SU(3) \times U(1)] \times SO(4)$, while
$\VEV{A^\prime}$
and $\VEV{S^\prime}$ break $SO(10) \rightarrow SO(6) \times [SU(2) \times
U(1) \times U(1)]$. Altogether, then, the unbroken group is $SU(3)
\times SU(2) \times U(1) \times U(1) =$ rank 5.

To break all the way to the standard model further Higgs fields are
needed. (They are needed for right-handed neutrino masses in any
case.) The simplest choices are $16 + \overline{16}$ or $126 +
\overline{126}$. Let us call these $C + \bar{C}$ where $C = 16$ or
126. One can write a superpotential that gives $C + \bar{C}$ VEVs
which break $SU(10) \rightarrow SU(5)$. Together with $A, S,
A^\prime$, and $S^\prime$ this will complete the breaking of $SO(10)$
to $G_S$ and allow $\nu_R$ masses.

At this point a further somewhat subtle technical problem arises.
There are certain generators of $SO(10)$ that are broken both by $C +
\bar{C}$ and by the adjoints $A$ and $A^\prime$, but not by the
symmetric tensors $S + S^\prime$, specifically the generators in
$SO(6)/SU(3) \times U(1)$ and $SO(4)/SU(2) \times U(1)$ (where
$SO(10) \supset SO(6) \times SO(4)$). Thus to avoid residual
goldstone bosons there must be coupling between the $C + \bar{C}$
sector and the adjoints $A$ and $A^\prime$. The technical problem is
that the direct couplings $\bar{C} A C$ and $\bar{C} A^\prime C$
would destabilize the desired VEVs of $A$ and $A^\prime$. In
particular, \underline{all} the `diagonal' components of $\VEV{A}$ and
$\VEV{A^\prime}$ (written as $U(5)$ matrices, that is) become
non-vanishing. [This is because the VEVs of $C + \bar{C}$ that break
$SO(10) \rightarrow SU(5)$ couple in $\bar{C}AC$ to the
$SU(5)$-singlet combination $(3a + 2b)$, where $\VEV{A} = \left(
\begin{array}{cc} 0 & 1\\ -1 & 0\end{array} \right) \times {\rm diag}
(a, a, a, b, b)$. This leads to a mass term proportional to $(3a +
2b)^2$ in the ordinary scalar potential, which in turn leads to
cross-terms of the form $a b$, destabilizing the solution $a \neq
0$ and $b = 0$. The same thing would happen to $A^\prime$.]

There are various solutions to this technical difficulty. The one we
shall study here involves the introduction of a third adjoint, which
we will denote $A^{\prime\prime}$, that serves as an intermediary
between the $C + \bar{C}$ sector and $A$ and $A^\prime$.

The part of the superpotential that does the complete breaking to the
standard model and obviates
\underline{all} difficulties is given in full by

\begin{equation}\begin{array}{lcl}
W & = & m_1 A^2 + m_2 S^2 + \lambda_1 S^3 + \lambda_2 A^2 S\strut\\
& + & m_1^\prime A^{\prime 2} + m_2^\prime S^{\prime 2} + \lambda_1^\prime
S^{\prime 3} + \lambda_2^\prime A^{\prime 2} S^{\prime}\strut\\
& + & m_1^{\prime\prime} A^{\prime\prime 2} + m_2^{\prime\prime}
\bar{C} C + \lambda_2^{\prime\prime} \bar{C} A^{\prime\prime}
C\strut\\
& + & \lambda A A^\prime A^{\prime\prime}.\end{array}
\end{equation}

\noindent
There are three sectors, $(A,S), (A^\prime, S^\prime)$, and
$(A^{\prime\prime}, \bar{C}+ C)$, that are coupled together only by
the last term $\lambda A A^\prime A^{\prime\prime}$.

The term $\lambda_2^{\prime\prime} \bar{C} A^{\prime\prime} C$ does
serve to give $A^{\prime\prime}$ a VEV in the $SU(5)$-singlet
direction:

\begin{equation}
\VEV{A^{\prime\prime}} = \left( \begin{array}{cc} 0 & 1\\ -1 &
0\end{array} \right) \otimes {\rm diag} (a^{\prime\prime},
a^{\prime\prime}, a^{\prime\prime}, a^{\prime\prime},
a^{\prime\prime}).
\end{equation}

\noindent
But this does not destabilize the VEVs of $A$ and $A^\prime$ which
are assumed to be of the forms given in eqs. (4) and (7). This is
easily seen by examining the $\lambda A A^\prime A^{\prime\prime}$
term, which is the only thing linking $A^{\prime\prime}$ to $A$ and
$A^\prime$. Consider the $F_A = 0$ equation. $(F_A)^{[ab]}$ is an
antisymmetric tensor to which $\lambda A A^\prime
A^{\prime\prime}$ contributes $\lambda (A^\prime
A^{\prime\prime})^{[ab]}$, which vanishes when the values of
$\VEV{A^\prime}$ and $\VEV{A^{\prime\prime}}$ given in eqs (7) and (18) are
substituted. Similarly, at the desired minimum, $\lambda A A^\prime
A^{\prime\prime}$ gives no contribution to the $F_{A^\prime}$ and
$F_{A^{\prime\prime}}$ equations. In other words it can be neglected
in doing the minimization! However, it does contribute to the
Higgs(ino) masses, and, indeed, removes all of the possible goldstone
modes discussed above.

In Appendix A we present details of the minimization of the
superpotential, eq. (17), assuming $C = 16$. There we show explicitly
that $SO(10)$ may be completely broken to the standard model (without
breaking SUSY), uneaten goldstone bosons avoided, and VEVs of the
desired form achieved. The masses of the various Higgs (super)
multiplets enumerated in Appendix A will be used to estimate
threshold corrections in the model.

It is conceivable that the superpotential of eq. (17) is the most
general one compatible with some discrete symmetries, although due to
the supersymmetric non-renormalization theorem one is not obliged to write
all possible terms.

We have found other superpotentials and sets of Higgs fields that
allow us to achieve the desired VEVs in a consistent and realistic
way. We have presented eq. (17) as being algebraically simple to
analyze. It should also be noted that the implementation of weak
suppression of HMPD, where only a single $45_H$ is needed with VEV of
the form in eq. (4), is a simpler task and fewer fields are required.
We have not tried to find the absolutely minimal scheme.

At this point we wish to make an aside. If one is willing to give up
$\sin^2\theta_W$ as a \underline{prediction}, there is a much simpler
way to simultaneously achieve doublet-triplet splitting without
fine-tuning and a strong suppression of
higgsino mediated proton
decay. All we need is then one $45_H$ of Higgs superfield with its
VEV as given in eq. (4). Suppose the relevant superpotential term for
doublet-triplet splitting is just $\lambda 10_{1H} 45_H 10_{2H}$ as
in eq. (5). This will make the color triplets heavy, but one is left
with 2 pairs of light doublets. Now, if the $45_H$ does not couple to
the sector that breaks $SO(10) \rightarrow SU(5)$ (via $C + \bar{C}$
superpotential), then in addition to the extra pair of doublets, one
will have a $\{ (3, 1, \frac{2}{3}) + H.C.\}$ goldstone
super-multiplet which remains light. (These are the goldstones
corresponding to $SO(6)/SU(3) \times U(1)$ mentioned earlier.) The
combined effect of having an extra pair of light higgs doublets and
the charge $\frac{2}{3}$ higgs (super) fields is to alter $\sin^2\theta_W$
prediction to $\simeq 0.215$ at one loop. The unification scale also
comes down by an order of magnitude or so. If one `fixes' these
features by relying on particle thresholds, such a scenario may not
be inconsistent. This scenario can be tested by directly searching
for the $(3, 1, \frac{2}{3})$ - higgs and higgsino particles.
This situation is somewhat analogous to the case studied in ref. 17.
We do not advocate this scenario
here, since our aim is to preserve the successful unification of
couplings as a prediction.

Returning to the superpotential in eq. (17), it might be imagined
that with $3(45) + 2(54) + \overline{16} + 16$ the threshold
corrections might be fairly large; large enough, perhaps to vitiate
the successful `prediction' of $\sin^2\theta_W$. Actually, this is
not the case, especially if one assumes $SO(10)$ breaks in two stages
to the standard model: $SO(10) \rightarrow SU(5) \rightarrow SU(3)
\times SU(2) \times U(1)$. $SO(10)$ is broken to $SU(5)$ by the VEVs
of $\bar{C}, C$, and $A^{\prime\prime}$ at a scale $M_{10}$, while
$SU(5)$ is broken down to $SU(3) \times SU(2) \times U(1)$ by the
VEVs of $A, S, A^\prime$, and $S^\prime$ at a scale $M_5$. The masses
of particles will be of the form $\alpha M_{10} + \beta M_5$. In the
limit that $\beta M_5/\alpha M_{10} \rightarrow 0$ for a given
multiplet its one-loop threshold corrections to $\sin^2\theta_W$ will
vanish since it will become a complete and degenerate $SU(5)$
multiplet. Thus threshold corrections of complete $SU(5)$-multiplets
go as $\ell n (\alpha M_{10} + \beta M_5)/(\alpha M_{10}) \cong \beta
M_5/\alpha M_{10}$ for $M_5 \ll M_{10}$. Thus if $M_{10}$ is assumed
to be somewhat larger than $M_5$, the GUT-scale threshold corrections
to $\sin^2\theta_W$ are substantially reduced. These will be
discussed explicitly in Appendix B, where it is found that the
uncertainties in $\sin^2\theta_W$ due to superheavy thresholds is
typically in the range of $3 \times 10^{-3}$ to $10^{-2}$.

\section*{6. Conclusions}

If one seeks a supersymmetric grand unified model in which proton
decay mediated by color-triplet higgsino is strongly suppressed
through a mechanism based on symmetry, in which there is no
fine-tuning of parameters, and in which the remarkable prediction of
$\sin^2\theta_W$ is maintained \underline{as a prediction}, then it
seems that one must turn to $SO(10)$. On the other hand, a weak
suppression due not to symmetry but to the smallness of a parameter
is achievable in both $SU(5)$ and $SO(10)$ without either fine-tuning
or sacrificing the $\sin^2\theta_W$ prediction, though we believe
$SO(10)$ allows the more economical solution. The $SU(5)$ solution,
being based on the `missing-partner mechanism', requires the
introduction of Higgs(ino) multiplets in $50 + \overline{50} + 75$
(which are four and five index tensors), whereas the $SO(10)$
solution requires only the usual (rank one and two) tensors 45, and
54, and the spinors $16 + \overline{16}$.

The advantage of $SO(10)$ is due to the possibility of exploiting the
elegant Dimopoulos-Wilczek mechanism of doublet-triplet splitting. We
have studied that mechanism and found that it can be implemented in a
fully realistic way.

In our view these results constitute yet another argument in favor of
$SO(10)$. It is already well known that $SO(10)$ has the advantage
over $SU(5)$ of allowing $R$-parity to be a gauge symmetry (that is
because Higgs fields are in tensor representations and matter fields
are in spinor representations). And, of course, $SO(10)$ achieves
greater unification of quarks and leptons and requires the existence
of right-handed neutrinos.

In any event, we have shown that higgsino-mediated proton decay is
not a serious difficulty of supersymmetric grand unification as there
are quite simple and natural means to suppress it without
undercutting the main success of those models. If the suppression is
of the `weak' type then there are grounds to hope to see proton decay
in super Kamiokande.

\def\VEV#1{\left\langle #1\right\rangle}

\def\theequation{\ksection.\arabic{equation}}
\def\ksection{A}              
\setcounter{equation}{0}
\newpage

\section*{Appendix A}

In this Appendix, we give details of the minimization of the
superpotential of eq. (17). We shall see explicitly that {\bf (a)} $SO(10)$
breaks completely to the standard model in the supersymmetric limit,
{\bf (b)} the desired forms of the VEVs of $A$ and $A^\prime$ are
achieved, and {\bf (c)} there are no unwanted pseudo goldstone modes
which could potentially ruin the successful $\sin^2\theta_W$
prediction.

$A^2$ in eq. (17) denotes $Tr (A^2),\; AA^\prime A^{\prime\prime}$
denotes $Tr (AA^\prime A^{\prime\prime})$ etc. We shall confine
ourselves to the case where $C+\bar{C} \equiv 16 + \bar{16}$. The
term $\bar{C}A^{\prime\prime}C$ is explicitly written down as
$\bar{C}{\sigma_{\alpha\beta}A^{\prime\prime}_{\alpha\beta}}
C/4$, where $\sigma_{\alpha\beta}$ are the generators of $SO(10)$
algebra.$^{18}$

The VEVs for the fields are chosen as

\begin{equation}\begin{array}{ccl}
\VEV{A} & = & \eta \otimes {\rm diag} (a,a,a,0,0) \strut\\
\VEV{A^\prime} & = & \eta \otimes {\rm diag} (0,0,0,a^\prime, a^\prime)
\strut\\
\VEV{A^{\prime\prime}} & = & \eta \otimes (a^{\prime\prime}, a^{\prime\prime},
a^{\prime\prime}, a^{\prime\prime}, a^{\prime\prime})
\strut\\
\VEV{S} & = & I \otimes {\rm diag} (s,s,s,-\frac{3}{2} s, -\frac{3}{2}s)
\strut\\
\VEV{S^\prime} & = & I \otimes {\rm diag} (s^\prime, s^\prime, s^\prime, -
\frac{3}{2} s^\prime, -\frac{3}{2} s^\prime) \strut\\
\VEV{C} & = & \VEV{\bar{C}}~ = ~c\end{array}
\end{equation}

\noindent
where $\eta \equiv \left( \begin{array}{cc} 0 & 1\\ -1 & 0\end{array}
\right)$ and $I = \left( \begin{array}{cc} 1 & 0\\ 0 & 1\end{array}
\right)$. The equality $\VEV{C} = \VEV{\bar{C}}$ follows from the requirement
of D-flatness.  The vanishing of the F--terms lead to the following
conditions, corresponding to
$A, S, A^\prime, S^\prime, A^{\prime\prime}, C$ fields
respectively:

\begin{eqnarray}
0 & = & m_1 + \lambda_2 s \nonumber\strut\\
0 & = & m_2 s - \frac{3}{4} \lambda_1 s^2 - \frac{1}{5} \lambda_2 a^2
\nonumber\strut\\
0 & = & m_1^\prime - \frac{3}{2} \lambda^\prime_2 s^\prime \nonumber\strut\\
0 & = & m_2^\prime s^{\prime} - \frac{3}{4} \lambda_1^\prime s^{\prime 2} +
\frac{1}{5} \lambda_2^\prime a^{\prime 2}  \nonumber\strut\\
0 & = & m_1^{\prime\prime} a^{\prime\prime} +
\frac{\lambda_2^{\prime\prime}}{8} c^2 \nonumber\strut\\
0 & = & m_2^{\prime\prime} - \frac{5}{2} \lambda_2^{\prime\prime}
a^{\prime\prime} ~.
\end{eqnarray}

Since SUSY is unbroken, it is sufficient to investigate the Higgsino
mass spectrum. The multiplets which transform as $\{ (3, 1, 2/3) +
H.C.\}$ under $SU(3)_C \times SU(2)_L \times U(1)_Y$ have the
following mass matrix:

\begin{equation}
{\cal M}_1 = \left( \begin{array}{cccc}
0 & 2\lambda a^{\prime\prime} & 0 & 0\strut\\
2\lambda a^{\prime\prime} & - 10 \lambda_2^\prime b^\prime &
-2\lambda a & 0\strut\\
0 & -2\lambda a & -4 m_1^{\prime\prime} & -\lambda_2^{\prime\prime}
c\strut\\
0 & 0 & -\lambda_2^{\prime\prime} c & 2\lambda_2^{\prime\prime}
a^{\prime\prime}\end{array} \right)~.
\end{equation}

\noindent
This matrix has one zero eigenvalue by virtue of eq. (A.2).
All the other 3 states become massive.

The mass matrix corresponding to $\{ (1,1, +1) + H.C. \}$ is

\begin{equation}
{\cal M}_2 = \left( \begin{array}{cccc}
10 \lambda_2 b & 2\lambda a^{\prime\prime} & -2\lambda a^\prime &
0\strut\\
2\lambda a^{\prime\prime} & 0 & 0 & 0\strut\\
- 2\lambda a^\prime & 0 & -4 m_1^{\prime\prime} &
-\lambda_2^{\prime\prime} c\strut\\
0 & 0 & -\lambda_2^{\prime\prime} c & 2\lambda_2^{\prime\prime}
a^{\prime\prime}\end{array} \right)~.
\end{equation}

\noindent
Again, ${\cal M}_2$ has one zero eigenvalue (using (A.2)) and
three non-zero eignevalues.

The mass matrix for $\{ (3,2,-\frac{5}{6}) + H.C.\}$ is given by

\begin{equation}
{\cal M}_3 = \left( \begin{array}{ccccc}
5 \lambda_2 s & \sqrt{2} \lambda_2 a & 0 & 0 & -\lambda
a^\prime\strut\\
\sqrt{2} \lambda_2 a & 2 m_2 -\frac{3}{2} \lambda_1 s & 0 & 0 &
0\strut\\
0 & 0 & -5 \lambda_2^\prime s^\prime & \sqrt{2} \lambda_2^\prime
a^\prime & -\lambda a\strut\\
0 & 0 & \sqrt{2} \lambda_2^\prime a^\prime & 2m_2^\prime -\frac{3}{2}
\lambda_1^\prime s^\prime & 0\strut\\
-\lambda a^\prime & 0 & -\lambda a & 0 & -4
m_1^{\prime\prime}\end{array} \right)~.
\end{equation}

\noindent
Using (A.2) one sees that ${\cal M}_3$ has one of its eigenvalues
equal to zero, while the rest are all nonzero.

The corresponding matrix for $\{ (3,2,\frac{1}{6}) + H.C.\}$ is given
by

\begin{equation}
{\cal M}_4 = \left( \begin{array}{cccccc}
5 \lambda_2 s & \sqrt{2} \lambda_2 a & 2 i \lambda a^{\prime\prime} &
0 & -i \lambda a^\prime & 0\strut\\
\sqrt{2} \lambda_2 a & 2 m_2 -\frac{3}{2} \lambda_1 s & 0 & 0 & 0 &
0\strut\\
-2 i \lambda a^{\prime\prime} & 0 & -5 \lambda_2^\prime b^\prime &
\sqrt{2} \lambda_2^\prime a^\prime & i \lambda a & 0\strut\\
0 & 0 & \sqrt{2} \lambda_2^\prime a^\prime & 2m_2^\prime -\frac{3}{2}
\lambda_1^\prime s^\prime & 0 & 0\strut\\
i \lambda a^\prime & 0 & -i \lambda a & 0 & -4m_1^{\prime\prime} &
\lambda_2^{\prime\prime} c\strut\\
0 & 0 & 0 & 0 & \lambda_2^{\prime\prime} c &
2\lambda_2^{\prime\prime} a^{\prime\prime}\end{array} \right)
\end{equation}

\noindent
This has one zero and five nonzero eigenvalues.

The goldstone modes in ${\cal M}_1, {\cal M}_2, {\cal M}_3,
{\cal M}_4$ when combined with the
zero mass mode corresponding to the phase of $(C + \bar{C})$--singlet
add up to the 33 mass--less modes needed for the symmetry breaking
$SO(10) \rightarrow SU(3) \times SU(2) \times U(1)$. All the
remaining fields become massive. Their spectrum looks as follows.
{}From the $(A,S)$ sector, we have

\begin{eqnarray}
\{ (6, 1, -\frac{2}{3}) + H.C.\} & = & 2 m_2 + 6 \lambda_1
s\nonumber\strut\\
\{ (1, 3, \pm 1);~ (1, 3, 0)\} & = & 2 m_2 - 9 \lambda_1
s\nonumber\strut\\
\{ (1, 3, 0) + (1, 1, 0) \} & = & 10 \lambda_2 s\nonumber\strut\\
\{ (8, 1, 0)\} & = & \left( \begin{array}{cc} 0 & 2 \sqrt{2} \lambda_2
a\strut\\
2 \sqrt{2} \lambda_2 a & 2 m_2 + 6 \lambda_1 s\end{array}
\right)\nonumber\strut\\
\{(1, 1, 0)\} & = & \left( \begin{array}{cc} 0 & -\frac{4}{\sqrt{5}}
\lambda_2 a\strut\\ -\frac{4}{\sqrt{5}} \lambda_2 a & 2 m_2 - 3\lambda_1
s\end{array} \right)~.
\end{eqnarray}

\noindent
{}From the $(A^\prime, S^\prime)$ sector, one finds

\begin{eqnarray}
\{ (6, 1, -\frac{2}{3}) + H.C.;~ (8, 1, 0) \} & = &  2 m_2^\prime + 6
\lambda_1 s^\prime\nonumber\strut\\
\{ (8, 1, 0);~ (1, 1, 0) \} & = & -4 \lambda_2^\prime
s^\prime\nonumber\strut\\
\{ (1, 3, \pm 1) \} & = & 2 m_2^\prime - 9 \lambda_1^\prime
s^\prime\nonumber\strut\\
\{ (1, 3, 0) \} & = & \left( \begin{array}{cc} 0 & -2 \sqrt{2}
\lambda_2^\prime a^\prime\strut\\
-2\sqrt{2} \lambda_2^\prime a^\prime & 2m_2^\prime - 9
\lambda_1^\prime s^\prime\end{array} \right)\nonumber\strut\\
\{ (1, 1, 0) \} & = & \left( \begin{array}{cc} 0 & \sqrt{\frac{24}{5}}
\lambda_2^\prime a^\prime\strut\\
\sqrt{\frac{24}{5}} \lambda_2^\prime a^\prime & 2 m_2^\prime - 3
\lambda_1^{\prime} s^\prime\end{array} \right)~.
\end{eqnarray}

\noindent
Finally, from the $(A^{\prime\prime}, C + \bar{C})$ sector, one finds

\begin{eqnarray}
\{ (8, 1, 0) + (1, 3, 0) + (1, 1, 0) \}  & = & -4
m_1^{\prime\prime}\nonumber\strut\\
\{ (3, 1, -\frac{1}{3} ) + H.C.;~ (1, 2, \frac{1}{2} ) + H.C. \} & = & 2
m_2^{\prime\prime} + 3 \lambda_2^{\prime\prime}
a^{\prime\prime}\nonumber\strut\\
\{(1, 1, 0)\} = \left( \begin{array}{cc} -4 m_1^{\prime\prime} & -
\sqrt{\frac{5}{2}} \lambda_2^{\prime\prime} c\strut\\
- \sqrt{\frac{5}{2}} \lambda_2^{\prime\prime} c & m_2^{\prime\prime}
- \frac{5}{2} \lambda_2^{\prime\prime} a^{\prime\prime}\end{array}
\right)
\end{eqnarray}

Now to reduce threshold corrections somewhat (and simplify
calculations) we assume the scale of $SO(10)$ breaking, $M_{10}$, is
somewhat greater than the scale of $SU(5)$ breaking, $M_5$. (This
means we assume $m_1^{\prime\prime}, m_2^{\prime\prime}, c, a^{\prime\prime}
\gg m_1, m_2, m_1^\prime, m_2^\prime, a, a^\prime, s, s^\prime$.) As
explained in the text, the multiplets which have mass $O (M_{10})$
will give contributions to the threshold corrections suppressed by
$M_5/M_{10}$. Thus in Appendix B we will only need the masses of
particles which are $O(M_5)$. For sets of particles with the same
$G_S$ quantum numbers we only will need to know the products of their
masses. These are listed below.

\begin{equation}\begin{array}{rcll}
\{ (3, 2, \frac{1}{6}) + h.c. \} & = & (\frac{4}{25})
\lambda_2\lambda_2^\prime a^2 a^{\prime 2}/s s^\prime & {\rm (2\
states)}\strut\\
\{ (3, 2, -\frac{5}{6}) + h.c.\} & = & (\frac{4}{25})
(\lambda^2\lambda_2\lambda_2^\prime/\lambda^{\prime\prime}(a^2a^\prime
a^{\prime\prime}/ss^\prime c^2) &\strut\\
& & \cdot [2aa^\prime (a + a^\prime) + 25 (a^\prime s^2 +
as^{\prime 2})] & {\rm (3\ states)}\strut\\
\{ (6, 1, -\frac{2}{3}) + h.c. \} & = & (\frac{15}{2}) \lambda_1 s +
(\frac{2}{5}) \lambda_2 a^2/s & {\rm (1\ state)}\strut\\
\{ (1, 3, \pm 1);\ (1, 3, 0)\} & = & -(\frac{15}{2}) \lambda_1 s +
(\frac{2}{5}) \lambda_2 a^2/s & {\rm (1\ state)}\strut\\
\{ (1, 3, 0)\} & = & 10\lambda_2 s & {\rm (1\ state)}\strut\\
\{ (8, 1, 0)\} & = & 8\lambda_2^2 a^2 & {\rm (2\  states)}\strut\\
\{ (6, 1, -\frac{2}{3}) + h.c.;\ (8, 1, 0)\} & = & (\frac{15}{2})
\lambda_1^\prime s^\prime - (\frac{2}{5}) \lambda_2^\prime a^{\prime
2}/s^\prime & {\rm (1\ state)}\strut\\
\{ (8, 1, 0)\} & = & -4 \lambda_2^\prime s^\prime & {\rm (1\
state)}\strut\\
\{ (1, 3, \pm 1)\} & = & -(\frac{15}{2}) \lambda_1^\prime s^\prime -
(\frac{2}{5}) \lambda_2^\prime a^{\prime 2}/s^\prime & {\rm (1\
state)}\strut\\
\{ (1, 3, 0)\} & = & 8 \lambda_2^{\prime 2} a^{\prime 2} & {\rm (2\
states)}\end{array}
\end{equation}

\noindent
We have expressed these in terms of the VEVs and Yukawa couplings and
eliminated the mass parameters, $m_1, m_2$, etc., using eqs. (A.2).

\newpage
\def\ksection{B}
\setcounter{equation}{0}
\section*{Appendix B}

Here we use the results of Appendix A to compute the threshold
corrections to $\sin^2\theta_W$ coming from superheavy fields. We
look at the full model with three 45's $(A, A^\prime,
A^{\prime\prime})$ and two 54's $(S, S^\prime)$, and we assume
$SO(10)$ breaks to $SU(5)$ at a scale $(M_{10})$  which is higher than
the scale at which $SU(5)$
breaks to the standard model $(M_5)$.

It will prove convenient to define the parameters $M \equiv \lambda_2
s,\ x \equiv a/s, x^\prime \equiv a^\prime/s^\prime,\ y \equiv
\lambda_1/\lambda_2,\ y^\prime \equiv
\lambda_1^\prime/\lambda_2^\prime$, and $z \equiv\lambda_2^\prime
s^\prime/\lambda_2 s$. The correction to $\sin^2\theta_W$ is given at
one loop by

\begin{equation}
\Delta \sin^2\theta_W (M_Z) = \frac{\alpha (M_Z)}{30\pi} \sum_j
\left[ 5 b_1^j - 12 b_2^j + 7 b_3^j \right] \ln\ M_j
\end{equation}

\noindent
where the sum is taken over multiplets of $SU(3) \times SU(2) \times
U(1)$. From eqs. (A.10) one obtains

\begin{equation}\begin{array}{rcl}
\Delta\sin^2\theta_W (M_Z) & = & \frac{\alpha (M_Z)}{30 \pi^2} \{ -
21 \ln (\frac{4}{25} x^2 x^{\prime 2} z M^2)\strut\\
& + & 3 \ln (t M^3) + 51 \ln ((\frac{15}{2} y + \frac{2}{5} x^2)
M)\strut\\
& + & (-30 - 24) \ln ((-\frac{15}{2} y + \frac{2}{5} x^2) M) - 24 \ln
(10 M)\strut\\
& + & 21 \ln (8 x^2 M^2) + (51 + 21) \ln ((\frac{15}{2} y^\prime -
\frac{2}{5} x^{\prime 2}) z M)\strut\\
& + & 21 \ln (4 z M) - 30 \ln (( -\frac{15}{2} y^\prime - \frac{2}{5}
x^{\prime 2} ) z M)\strut\\
& - & 24 \ln (8 x^{\prime 2} z^2 M^2) \}.\end{array}
\end{equation}

\noindent(Here $tM^3$ is defined to be equal to the complicated
expression on the right-hand side of the second equation of (A.10).
$t$ is of order $M_5/M_{10}$. However, the coefficient of this term
is mercifully small, so that its effect is negligible.  We have not
displayed the threshold effects due to the doublet and triplet fileds of
eqs. (9)-(10) as they are negligible.)  All logarithms are
understood to have an absolute value in their arguments.  We have
used $b_3(3) = \frac{1}{2},\ b_3(6) = \frac{5}{2},\ b_3(8) = 3,\
b_2(2) = \frac{1}{2},\ b_2(3) = 2,\ b_1(\frac{y}{2}) = \frac{3}{5}
(y/2)^2$.

Collecting terms,

\begin{equation}\begin{array}{rcl}
\Delta\sin^2\theta_W(M_Z) & \cong & \frac{\alpha(M_Z)}{30\pi} \{
(18\ln 5 - 33 \ln 2)\strut\\
& + & 3\ln t - 6 \ln z - 45 \ln (x^{\prime 2})\strut\\
& + & 51 \ln (\frac{15}{2} y + \frac{2}{5} x^2) - 54 \ln
(\frac{15}{2} y - \frac{2}{5} x^2)\strut\\
& + & 72 \ln (\frac{15}{2} y^\prime - \frac{2}{5} x^{\prime 2}) - 30
\ln (\frac{15}{2} y^\prime + \frac{2}{5} x^{\prime 2})\strut\\
& - & 3 \ln M \}\end{array}~.
\end{equation}

\noindent
The term $-3 \ln M$ is due to the $(3, 2, -\frac{5}{6}) + h.c.$ that
get eaten when $SU(5)$ breaks and is present also in the minimal
$SU(5)$ model. If it were not for this eating, all the superheavy
multiplets would be complete $SU(5)$ multiplets and the dependence on
$M$ would disappear. (It is only splitting within multiplets that
contribute at one loop to $\sin^2\theta_W$. Since all masses are
scaled by $M, M$ drops out in the ratios.)  In the first term,
$18 \ln 5 - 33 \ln 2 \cong
6.1$. All of the logarithms can have either sign. There are five
potentially large terms with coefficients averaging about 50. If we
assume the logarithms are of order one with arbitrary signs then the
typical threshold correction to $\sin^2\theta_W$ would be expected to
be about $\pm (\alpha/30\pi) (10^2) \sim \pm 10^{-2}$. This is to be
contrasted with the effect of a pair of extra light Higgs doublets of
$\Delta\sin^2\theta_W \simeq 2.5 \times 10^{-2}$. It should also be
compared to the theoretical uncertainties in $\sin^2\theta_W$ in
minimal SUSY $SU(5)$, referred to in the opening paragraph of this
paper, of about $0.36 \times 10^{-2}$.

The expression in (B.2) simplifies considerably if we assume $y <
\frac{4}{75} x^2,\ y^\prime < \frac{4}{75} x^{\prime 2}$.  Then

\begin{equation}\begin{array}{rcl}
\Delta\sin^2\theta_W & \simeq & \left( \frac{\alpha}{30 \pi} \right) \{ 3
\ln\ t - 3 \ln (x^2) - 3 \ln (x^{\prime 2}) \strut\\
& - & 6 \ln\ z + (6 \ln 2 - 21 \ln 5) \}\end{array}~.
\end{equation}

\noindent
Since $6 \ln 2 - 21 \ln 5 = - 29.4$ one expects the threshold
correction to be negative and about $-0.3 \times 10^{-2}$ in this
limit.  We mention this limit since it is a special solution of the
superpotential of eq. (17), corresponding to setting the parameters
$\lambda_1,~\lambda_1^\prime$ to zero.  We note that this limit can be
reached naturally without giving rise to any pseudo goldstones (see
Appendix A).

We conclude that the threshold correction uncertainties to
$\sin^2\theta_W$ in the kind of $SO(10)$ model we are discussing are
likely to be a few times larger than the total theoretical
uncertainty in $\sin^2\theta_W$ in minimal SUSY $SU(5)$, but several
times smaller than the effect on $\sin^2\theta_W$ of an extra pair of
light Higgs doublets. We should emphasize that if one is satisfied
only to suppress Higgsino mediated proton decay \underline{weakly}, a
much smaller Higgs sector may be adequate, with correspondingly
smaller threshold corrections.
\newpage

\section*{References}

\begin{enumerate}
\item U. Amaldi, W. de Boer, and H. F\"{u}rstenau, Phys. Lett.
\underline{260B}, 447 (1991);\\
P. Langacker and M.-X. Luo, Phys. Rev. \underline{D44}, 817 (1991);\\
J. Ellis, S. Kelley and D.V. Nanopoulos, Phys. Lett. \underline{260B},
131 (1991).
\item P. Langacker and N. Polonsky, Pennsylvania Preprint UPR-0513T
(1993).
\item N. Sakai and T. Yanagida, Nucl. Phys. \underline{197B}, 533
(1982);\\
S. Weinberg, Phys. Rev. \underline{D26}, 287 (1982).
\item H. Hisano, H. Murayama, and T. Yanagida, Tohoku University
preprint TU-400 July (1992).
\item J. Ellis, D.V. Nanopoulos and S. Rudaz, Nucl. Phys.
\underline{B202}, 43 (1982); \newline
P. Nath, A.H. Chamseddine and R. Arnowitt, Phys. Rev. \underline{D32},
2348 (1985); \newline
P. Nath and R. Arnowitt, Phys. Rev. \underline{D38}, 1479 (1988) and
Northeastern preprints NUB-TH-3041-92,
3045-92, 3046-92 (1992).
\item Kamiokande Collaboration, K.S. Hirata et. al., Phys. Lett.
\underline{B220}, 308 (1989).
\item G.D. Coughlan, G.G. Ross, R. Holman, P. Ramond, M. Ruiz-Altaba,
and J.W.F. Valle, Phys. Lett. \underline{158B}, 401 (1985).
\item J. Hisano, H. Murayama, and T. Yanagida, Phys. Lett.
\underline{291B}, 263 (1992).
\item A. Buras, J. Ellis, M. Gaillard, D.V. Nanopoulos, Nucl. Phys.
\underline{B135}, 66 (1985).
\item H. Georgi, Phys. Lett. \underline{108B}, 283 (1982);\\
A. Masiero, D.V. Nanopoulos, K. Tamvakis, and T. Yanagida, Phys.
Lett. \underline{115B}, 380 (1982);\\
B. Grinstein, Nucl. Phys. \underline{B206}, 387 (1982).
\item E. Witten, Phys. Lett. \underline{105B}, 267 (1981);\\
D.V. Nanopoulos and K. Tamvakis, Phys. Lett. \underline{113B}, 151
(1982);\\
S. Dimopoulos and H. Georgi, Phys. Lett. \underline{117B}, 287
(1982);\\
L. Ibanez and G. Ross, Phys. Lett. \underline{110B}, 215 (1982);\\
A. Sen, FERMILAB-PUB-84/67.
\item J. Polchinski and L. Susskind, Phys. Rev. \underline{D26},
3661 (1982);\\
M. Dine, lectures at Johns Hopkins Workshop (1982);\\
H.P. Nilles, M. Srednicki, and D. Wyler, Phys. Lett.
\underline{124B}, 237 (1982);\\
A.B. Lahanas, Phys. Lett. \underline{124B}, 341 (1982).
\item S. Dimopoulos and F. Wilczek, NSF-ITP-82-07 (unpublished).
\item M. Srednicki, Nucl. Phys. \underline{B202}, 327 (1982).
\item A. De Rujula, H. Georgi and S.L. Glashow, Phys. Rev. Lett.
\underline{45}, 413 (1980);\\
H. Georgi, S.L. Glashow and M. Machacek, Phys. Rev. \underline{D23},
783 (1981);\\
S.M. Barr, Phys. Lett. \underline{112B}, 219 (1982).
\item I. Antoniades, J. Ellis, J. Hagelin, and D.V. Nanopoulos, Phys.
Lett. \underline{194B}, 231 (1987);
ibid., \underline{205B}, 459 (1988);
ibid., \underline{208B}, 209 (1988);
ibid., \underline{231B}, 65 (1989).
\item G.R. Dvali, Phys. Lett. \underline{B287}, 101 (1992).
\item H. Georgi and D.V. Nanopoulos, Nucl. Phys. \underline{B 155}, 52
(1979).
\end{enumerate}

\newpage

\section*{Figure Captions}

\begin{enumerate}
\item A diagram that gives a dimension 5 baryon-number-violating
operator. The arrows indicate the direction a left-handed chiral
superfield is flowing. Chirality shows this to be an $F$ term and
hence suppressed by (Mass)$^{-1}$. Chirality also implies that there
must be a mass insertion (denoted by $M$) on the colored higgsino
line. The suppression is thus $M/M_{GUT}^2$ which is naturally of
order $M_{GUT}^{-1}$ in most models.

\item A diagram without the chirality-flipping mass insertion of Fig.
1 and thus representing a $D$ term. Such a term is effectively of
dimension 6 and suppressed by $M_{GUT}^{-2}$. It corresponds to
colored higgs boson exchange.

\item In the `weak suppression' scheme discussed in section I (see
eq. (2)) the primed and unprimed Higgs(ino) sectors are connected by
a Dirac mass, $M$. This coupling of the two sectors allows a
dimension-5, baryon-number-violating operator to arise as shown here.
\end{enumerate}

\end{document}